\documentstyle[12pt]{article}
\def\beq{\begin{equation}}   \def\eeq{\end{equation}}

\newcommand{\gsim}{\lower.7ex\hbox{$
\;\stackrel{\textstyle>}{\sim}\;$}}
\newcommand{\lsim}{\lower.7ex\hbox{$
\;\stackrel{\textstyle<}{\sim}\;$}}

\newcommand{\ra}{\rightarrow}

\newcommand{\La}{\overline{\Lambda}}
\newcommand{\Lam}{\Lambda_{\rm QCD}}

\renewcommand{\Im}{\mbox {Im}\:}

\newcommand{\be}{\beta}

\newcommand{\de}{\delta}
\newcommand{\al}{\alpha}
\newcommand{\as}{\alpha_s}

\newcommand{\aw}{\alpha_s^{(\omega)} }
\newcommand{\am}{\alpha_s^{\overline{\rm MS}} }

\newcommand{\GeV}{\,\mbox{GeV}}

\newcommand{\matel}[3]{\langle #1|#2|#3\rangle}

\newcommand{\aver}[1]{\langle #1 \rangle}

\begin{document}

\def\lsim{\mathrel{\rlap{\lower3pt\hbox{\hskip0pt$\sim$}}
    \raise1pt\hbox{$<$}}}         %less than or approx. symbol
\def\gsim{\mathrel{\rlap{\lower4pt\hbox{\hskip1pt$\sim$}}
    \raise1pt\hbox{$>$}}}         %greater than or approx. symbol

\begin{titlepage}
\renewcommand{\thefootnote}{\fnsymbol{footnote}}

\begin{flushright}
UND-HEP-97-BIG\hspace*{.2em}06\\
TTP97-33\\
hep-ph/9708372\\
\end{flushright}
\vspace{.3cm}
\begin{center} \Large
{\bf Non-Abelian Dipole Radiation and the\\ Heavy Quark Expansion}
\end{center}
\vspace*{.3cm}
\begin{center} {\Large
A.~Czarnecki $^{a}$, K.~Melnikov $^{a}$ and N. Uraltsev $^{b,c}$\\
\vspace{.4cm}
{\normalsize
$^a${\it Institut f\"ur Theoretische Teilchenphysik, \\
Universit\"at Karlsruhe,
D-76128 Karlsruhe, Germany}\\
$^b$ {\it Dept. of Physics, Univ. of Notre Dame du Lac, Notre Dame, IN 46556, 
U.S.A.\\
$^c$ {\it St.\,Petersburg Nuclear Physics Institute,
Gatchina, St.\,Petersburg 188350, Russia\footnote{Permanent address}
}}\\
}}
\vspace*{1.8cm}

{\Large{\bf Abstract}}\\
\end{center}
Dipole radiation in QCD is derived to the  second order in $\alpha_s$.
A power-like evolution of the spin-singlet heavy quark operators
is obtained to the same accuracy. In particular,  
${\cal O}(\alpha_s^2)$ relation between a short-distance low-scale 
running heavy quark mass and the $\overline{\rm MS}$ mass is given. 
We discuss the properties of the effective QCD coupling $\aw(E)$ which 
governs the dipole radiation. This coupling 
is advantageous for heavy quark physics.

\vspace*{.2cm}
\vfill
\noindent
\end{titlepage}
\addtocounter{footnote}{-1}

\newpage

Theoretical description of heavy flavor decays 
benefits from a strong hierarchy between the mass  of the
decaying quark  and the typical scale of the strong
interactions, $m_{b(c)} \gg \Lam$. 
A current level of experimental precision requires
an accurate
treatment of nonperturbative effects even in beauty decays. 
A consistent
genuinely QCD-based framework for such a treatment 
is provided by the heavy quark
expansion (HQE). The HQE  combines the Wilson operator product expansion 
(OPE) separating the 
physics of low and high momentum scales with the nonrelativistic
expansion, to treat the nonperturbative effects originating at large
distances. It allows a 
simultaneous precision account for both
perturbative and nonperturbative effects.

An important class of applications of the HQE are semileptonic
weak transitions between $b$ and $c$ quarks, i.e. 
the situation 
when both initial and final
quarks are heavy. In particular, the most informative predictions can be
made for the transition amplitudes in the so-called small velocity (SV)
limit \cite{SV}, when the velocities of 
heavy
hadrons in
initial and final state 
are small. The heavy quark in this case plays a role of a
static (or slowly moving) source of the color Coulomb field which
affects the light degrees of freedom in the hadrons. This physical picture 
is formalized in the Wilsonian
field-theory approach by integrating out high-momentum degrees of freedom
of full QCD and resorting to an effective low-energy theory, where
highly excited hadronic states are not present.

A  peculiarity of the effective theory for heavy flavor
hadrons is its essentially Minkowskian nature. 
In QCD any process in which the
velocity of the heavy quark $Q$ changes, involves actual gluon radiation 
with energy and momentum in the whole range up to the 
quark masses. On the other hand, effective theories are called upon to
eliminate all high-momentum physics. Such a problem does not arise in
usual effective theories (typically formulated in the Euclidean space)
where all short-distance physics is virtual.
This necessitates a careful control over the radiation effects.

On the other hand,  when the energy loss is small compared to
$m_Q$, the radiation off the heavy colored particles is
almost a classical effect, 
and therefore a universal process-independent description is possible.
In the limit of small velocity it is a familiar dipole radiation. It
has, of course, some peculiarities in the non-Abelian theories like QCD.

Even though the dipole radiation has the most obvious manifestation in
the SV processes, it is relevant in a more general context,  for
example, for zero-recoil processes or inclusive decay widths.  The
reason is that the OPE ensures that the effects of soft physics
originating at the momentum scale well below $m_Q$  enter all genuinely
short-distance observables in a universal way, via its contribution to
the local heavy-quark operators.  The knowledge of the perturbative
dipole radiation allows one to determine the short-distance evolution of
a number of composite operators in the effective theory of heavy quarks.

In the present paper, we derive the non-Abelian dipole radiation to  
the second order in $\alpha_s$ and use this result to obtain a power 
evolution of a number of
spin-singlet heavy quark operators to  order $\alpha_s^2$. In particular,
a gauge-invariant relation between a short--distance low--scale running heavy
quark mass 
and the $\overline{\rm MS}$ mass $\bar{m}_Q(m_Q)$ is given, which becomes
a practical necessity since the two-loop accuracy in the processes with
heavy quarks is becoming the state of the art.

The radiation by the heavy charged particle occurring when its velocity 
changes
is a well-known effect from classical electrodynamics. It is most simply 
obtained by a direct computation of the Lienard--Wiechert retarded 
potentials $A_\mu(\vec r,\,t)$ at $\vec{r} \ra \infty$, and has the
form 
\beq
\frac{1}{\omega}\frac{{\rm d}I(\omega)}{{\rm d}\omega} \;=\; 
\frac{\alpha}{\pi} 
\left(\frac{1}{|\vec v\,|}\ln{\frac{1+|\vec v\,|}{1-|\vec v\,|} } -2 \right) 
\frac{1}{\omega}
\:=\: \frac{2}{3}\,\frac{\alpha}{\pi} 
\frac{\vec{v}^{\,2}}{\omega} \;+\;{\cal O}(\vec{v}^{\,4})\;,
\label{5}
\eeq
where $\omega$ is the radiated energy, $I(\omega)$ is intensity 
and $\alpha$ is the fine structure constant. 
We will be interested only in 
the dipole term $\propto \vec{v}^{\,2}$ in QCD, and 
do not consider
multipole radiation proportional to higher powers of $\vec v$.

In quantum electrodynamics the same relation holds, with 
$1/\omega\; {\rm d}I/{\rm d}\omega$ giving the
probability to radiate soft photon(s) with energy $\omega$. Moreover,
there are no higher-order corrections in $\al_{\rm em}$, provided $\omega
\ll m$ and the recoil effects of the radiating particle $m$
are neglected. Nontrivial corrections are suppressed by powers of $\omega/m$. 

A certain type of
corrections emerge    
only due to effects of polarization of the quantum vacuum 
by dipole radiation. It is, therefore, directly related to the 
running of the 
gauge coupling in the quantum field theory. The radiation of the real
(on-shell) photons is proportional to 
the physical value of the fine structure constant 
$\alpha\equiv \alpha_{\rm em}(0)$. 
However, because of the pair production,
the radiated energy is governed by the running coupling $\alpha(\omega)$.
For small $\omega$, all vacuum polarization effects are suppressed by powers 
of $\omega^2/m_e^2$ where $m_e$ is the lightest charged particle. 
Therefore, the soft radiation in QED is defined by $\alpha(0)$ without 
any correction, in accord with classical electrodynamics. This is expected 
since semiclassical approximation is parametrically justified in the limit
$\omega \ra 0$. 

The dipole radiation in the non-Abelian theory is quite different. 
Because of gluon self-interaction, the effective coupling increases
even in the absence of light flavors (it also becomes different for
different multipoles). Eventually it grows up to values
of order $1$ and the theory enters the strong interaction phase. At such
energies nonperturbative effects emerge which generate color confinement
shaping the spectrum of observed hadrons. 

Our main interest lies in the domain where
$\omega$ is large compared to $\Lam$ (the nonperturbative multipole 
expansion of the color-singlet systems was discussed in 
\cite{mult}). In this regime, perturbative calculations 
can be performed.
Below we introduce a few standard
notations used for heavy quarks.

Let us consider a general process of scattering of 
an external color-singlet
weak current $J$ with momentum $q$ on a heavy quark $Q$ in the SV  
kinematics. 
For simplicity, the initial quark is assumed to
be at rest.
The initial $Q$ and final state $\tilde Q$ quarks can have arbitrary masses;
however, both masses must  be large, 
so that the nonrelativistic expansion can be applied.
The SV limit 
$
\vec{v}\;=\; {\vec q}/{m_{\tilde Q}}\,, \;\; |\vec{v}\,| \ll 1
$
is kept by adjusting $\vec{q}$ appropriately.

The current $J$ must have a non-vanishing tree level
nonrelativistic limit, otherwise it can be arbitrary.
The most familiar cases are the time-like component of the vector current
$J_0^\dagger =\bar{Q} \gamma_0 \tilde Q$ or the scalar current 
$J_S^\dagger =\bar{Q} \tilde Q$.

The inclusive processes of the scattering on heavy quarks are described
by the corresponding structure function $w(q_0,\,\vec{q}\,)$ which 
is a sum of all 
transition probabilities induced by $J$ into the final states with
momentum $\vec{q}$ and energy  $m_Q +q_0$. 
Using the optical theorem, one represents the structure function $w$ as 
a discontinuity of the forward transition amplitude 
$T(q_0,\,\vec{q}\,)$ at physical values of $q_0$: 
\beq
T(q_0,\,\vec{q}\,)\;=\; \frac{i}{2m_Q} 
\int\, d^4 x\: {\rm e}\,^{-iqx}\; 
\matel{Q}{T\,J(x) J^\dagger (0)}{Q}\;,\;\;
w(q_0,\,\vec{q}\,) = 2\: \Im T(q_0,\,\vec{q}\,)\;.
\label{9} 
\eeq
The causality ensures that $T(q_0,\,\vec{q}\,)$ is an analytic function of 
$q_0$ in the upper half-plane.
In the heavy quark limit the spin degrees of freedom become irrelevant, and
Eqs.~(\ref{9}) assume averaging over spin 
states
(also color, for perturbative
calculations, etc.).

The structure functions have a threshold $q_0^{\rm min}$ corresponding to the
elastic transition:
\beq
q_0^{\rm min}\;= \; \sqrt{\vec{q}^{\,2}+m_{\tilde Q}^2}-m_Q \simeq 
m_{\tilde Q}-m_Q + \frac{m_{\tilde Q} \vec{v}^{\,2}}{2}\;.
\label{10}
\eeq
The variable $\omega$ measures the
hadron excitation energy in the final state:
\beq
\omega \;=\; q_0-q_0^{\rm min}\; \simeq \;
q_0 + m_Q-m_{\tilde Q} - \frac{m_{\tilde Q} \vec{v}^{\,2}}{2}\;.
\label{12}
\eeq
For simplicity, we  consider below the case of equal
masses, $m_{\tilde Q}=m_Q$, although nothing depends
on this assumption.

The nonrelativistic expansion implies that $\omega \ll m_Q$. On the
other hand, the perturbative treatment is justified at $\omega \gg
\Lam$, and this hierarchy will be assumed in what follows. The
nonperturbative aspects will be addressed afterwards. Resorting to the
perturbative calculations, we can use the quark states instead of actual
heavy hadrons, which was tacitly assumed above.

The structure function $w$ takes the following form in the heavy quark
limit:
\beq
w(\omega,\,\vec{v})\;=\; N\,\de(\omega) + \,\frac{2\vec{v}^{\,2}}{3}\, 
\frac{d(\omega)}{\omega} + {\cal O}\left(\vec{v}^{\,4}\right)\;.
\label{14}
\eeq
At $\vec v =0$ only the elastic peak is present. The excitations,
described by the second term, appear at the level $\sim 1/m_Q^2$ or
$\vec{v}^{\,2}$ \cite{SV}. 
The dipole radiation is described by the function $d(\omega)$. 

Motivated by the dipole radiation in QED, we define a coupling 
$\aw$ by projecting Eq. (\ref {14}) on its second term:
\beq
C_F \frac{\aw(\omega)}{\pi \omega}\;=\;
\lim_{\vec v \ra 0} \; \lim_{m_Q \ra \infty}\,\frac{3}{2\vec{v}^{\,2}}
\frac{w(\omega,\,\vec{v})}{\int_0^{\omega}\, 
w(\omega',\,\vec{v})\,{\rm d}\omega'}
\;.
\label{16}
\eeq
The denominator in the last ratio is necessary to get rid of the
overall normalization of the effective
nonrelativistic current (renormalized down to the momentum scale
near $\omega$). In this form the heavy quark limit $m_Q\ra \infty$
yields a finite result. It is important that the normalization integral
includes also the elastic peak, which makes it infrared safe at
arbitrary $\vec v$. 
On the other hand, the exact upper limit does not matter in the SV
kinematics, since it affects the ratio only by corrections $\sim
\vec{v}^{\,2}$.

Therefore, 
the inelastic structure function in the SV limit determines the
effective QCD running coupling driving the dipole radiation of gluons.
It is a dimensionless quantity, and thus is
a function of the ratio $\Lam/\omega$. The explicit coefficient in
Eq.~(\ref{16}) is adjusted to make $\aw$ equal to 
the bare coupling in the Born approximation.

It is important that the OPE and factorization of the infrared 
effects\footnote{Physically, it is merely existence of the
effective low-energy theory of heavy quarks.} ensures that $\aw(\omega)$
is universal, in the sense that:\\
a) it does not depend on the choice of the weak current $J$ probing the
heavy quark structure, or on the ratio of the quark masses;\\
b) as long as the onset of the quark-hadron duality is passed, there is
no dependence on the particular type of the initial heavy hadron. This
property always holds in the perturbative analysis.

The effective coupling $\aw$ obeys the usual renormalization group
evolution 
\beq
\omega \frac{{\rm d}}{{\rm d}\omega} \frac{\aw(\omega)}{\pi}
\; =\; -\beta\left(\frac{\aw(\omega)}{\pi} \right) = 
-\frac{\beta_0}{2}\left(\frac{\aw(\omega)}{\pi} \right)^2 - 
\frac{\beta_1}{8}\left(\frac{\aw(\omega)}{\pi} \right)^3-\,...\;\;,
\label{18}
\eeq
$$
\beta_0=\frac{11}{3}C_A-\frac{2}{3}n_f\,,\qquad
\beta_1=102-\frac{38}{3} n_f\;\;.
$$
The coefficients $\beta_2$ and higher are scheme-dependent and therefore
differ from the standard ones obtained for the $\overline{\rm MS}$ coupling.
Note that because of its physical definition, the
evolution of $\aw$ is properly defined when 
heavy flavor thresholds are passed.

Although the coupling $\aw$ is most natural for heavy quark
decays, in practice one needs to know its relation to the standard
reference coupling $\am$. We calculated it to the second order in
$\as^2$ using the technique of the Ref.~\cite{cmtech}:
$$
\frac{\aw(\mu)}{\pi}\;=\;
\frac{\am(\mu)}{\pi} + 
\left[\left(\frac{5}{3} - \ln{2}\right)\frac{\beta_0}{2}
- C_A \left(\frac{\pi^2}{6}-\frac{13}{12} \right)\right]
\left(\frac{\as}{\pi}\right)^2 \,+\,{\cal O}(\as^3)\;=
$$
\beq
=\; \frac{\am\left({\rm e}^{-5/3+\ln{2}} \mu\right)}{\pi}
- C_A \left(\frac{\pi^2}{6} -\frac{13}{12} \right)
\left(\frac{\as}{\pi}\right)^2 \,+\,{\cal O}(\as^3)\;,
\label{20}
\eeq
where $C_A=N_c$ for the ${\rm SU}(N_c)$ gauge group.

In the latter form
we absorbed the effect of running $\as$ into the redefinition of the
scale at which the $\overline{\rm MS}$ coupling is evaluated 
\cite{BLM}. The term $\sim
C_A$ represents the so-called genuine (non-BLM) second-order effect; it has
purely non-Abelian origin. We find that its coefficient is not
particularly large, $-1.685$; it is not small either as it happens in a
number of other observables. Let us note that a similar effective
coupling $\alpha_{\rm eff}$ governing emission of soft gluons in the  
ultrarelativistic case $v\ra 1$ has been also calculated to two loops 
\cite{v1}: 
\beq
\frac{\alpha_{\rm eff}(\mu)}{\pi} \;=\; \frac{\am(\mu)}{\pi} +
\left[\frac{5}{6}\frac{\beta_0}{2} -
C_A\left(\frac{\pi^2}{12} -\frac{1}{3} \right)\right]
\left(\frac{\as}{\pi}\right)^2
\;;
\label{aldok}
\eeq
its exact definition, however, is less direct and more complicated.
It differs from $\aw$ in the 
second-order terms, although the difference is small numerically. 

The BLM-type terms which account for the effects of running of the QCD
coupling in the leading-order calculations can be easily computed to
any order in this case.\footnote{All necessary expressions are given in
\cite{blmope}; $c_\pi$ in Eq.~(28) is just $C_F\cdot\aw(\mu)/\pi$ in the BLM
approximation. The BLM series has a finite radius of convergence,
$4/\beta_0 \left(1+\frac{25}{9\pi^2}\right)^{-1/2}$ in terms of $\am(\mu)$, or
$4/\beta_0$ in terms of $\am(\mu {\rm e}^{-5/6})$.}
They contain corrections $~\sim \left(\beta_0\as /\pi \right)^k$
where the factor $\beta_0$ is singled out by the dependence on the
number of light flavors \cite{beta0}. 
For example, the BLM estimate of the $\as^3$ term in the last
Eq.~(\ref{20}) is 
\beq
{\cal O}(\as^3)
\;\stackrel{{\rm BLM}}{=} \; -\left(\frac{\pi^2}{6}-\frac{31}{36}\right)
\left(\frac{\beta_0}{2}\right)^2 \left(\frac{\as}{\pi} \right)^3\;, \qquad 
\frac{\pi^2}{6}-\frac{31}{36}\simeq 0.78\;.
\label{22}
\eeq

For book-keeping purposes we also give here the massive quark ${\cal
O} (\as)^2$ contribution to $\aw$: 
with an additional massive quark $q$ the extra term 
in the last Eq.~(\ref{20}) is given by
\beq
-\frac{1}{6}\left(\frac{\as}{\pi}\right)^2 \left\{
\ln{x} - 4\ln{2}+ \frac{10}{3} +
 \vartheta(1-x)\, \int_x^1 \,\frac{dt}{t}\,
\left(1+\frac{t}{2}\right) \sqrt{1-t} \left(1+\frac{x}{2t}\right)
\sqrt{1-\frac{x}{t} } \right\},
\label{26v}
\eeq
where $x=4m_q^2/\mu^2$ and $\am(\mu)$ is assumed to be defined 
for $n_f+1$ light flavor.

A construction of the HQE in QCD requires an accurate definition of the
basic objects of the effective theory, in the first place the heavy
quark mass, kinetic operator and all other composite operators.  Being
defined consistently,  all they depend on normalization point. A
normalization-point--independent  pole mass of the heavy quark, although
appearing in the purely perturbative calculations at a given order,
cannot be completely defined at the level when the nonperturbative
effects are addressed. By the same token, the parameter $\La$ measuring
the difference between the heavy hadron mass and $m_Q$ in the heavy
quark limit, suffers from a similar uncertainty. Instead, one must use
the short-distance  mass $m_Q(\mu)$ and, correspondingly, $\La(\mu)=
\lim _{m_Q\ra \infty} M_{H_Q}-m_Q(\mu)$ with $\Lam \ll \mu \ll m_Q$
\cite{pole}. The low-scale short distance mass can be defined in 
different ways. We recall, however, that the standard 
${\overline {\rm MS}}$
mass cannot be used for $\mu \ll m_Q$
\cite{rev}. 
Problems, similar to those in the pole mass, also emerge in attempts 
to define spin-singlet operators without a powerlike mixing with
lower-dimension operators, including the unit operator.

The physically appropriate gauge-invariant 
scheme was suggested in \cite{blmope,five,rev} and is based on the SV
sum rules relating the moments of the SV structure functions to the
local heavy quark operators. The normalization point is introduced as an
upper cutoff in the integral over excitation energy. 
For example, for a heavy hadron $H_Q$ one defines 
\beq
\La(\mu)\; \equiv \; \lim_{m_Q \ra \infty}\;
\left[ M_{H_Q}-m_Q(\mu)\right]\, =
\, \lim_{\vec v \ra 0}\;
\lim_{m_Q \ra \infty}\, \frac{2}{\vec v^{\,2}} 
\frac{\int_0^\mu \; {\omega\, w(\omega,\,\vec v\,)\:{\rm d}\omega}} 
{\int_0^\mu \;{w(\omega,\, \vec v\,)\:{\rm d}\omega}} \;\;,
\label{27}
\eeq
\beq
\mu_\pi^2(\mu)\;\equiv \; \frac{\matel{H_Q}{\bar Q (i\vec{D}\,)^2 Q}{H_Q}_\mu}
{\matel{H_Q}{\bar Q Q}{H_Q}_\mu}
\,=\,
\lim_{\vec v \ra 0} \; \lim_{m_Q \ra \infty}\,\frac{3}{\vec{v}^{\,2}}
\frac{\int_0^{\mu}\,\omega^2\,  w(\omega,\,\vec{v})\:{\rm d}\omega}
{\int_0^{\mu}\:w(\omega,\,\vec{v})\:{\rm d}\omega} 
\;,
\label{28}
\eeq
etc. 
The above operator relations for the SV moments 
of the structure function suggest that the
perturbative evolution must be driven by one and the same function.
With the above  definitions
it is directly $\aw(\mu)$. From Eqs. (\ref {27},\ref{28}) one finds:
\beq
\frac{{\rm d}\La(\mu)}{{\rm d}\mu}\;=\;\frac {4}{3} \,C_F \frac {\aw
(\mu)}{\pi}\;,\qquad
\frac{{\rm d}\mu_\pi^2(\mu)}{{\rm d}\mu^2}\;=\;
C_F \frac{\aw(\mu)}{\pi}\;,
\label{30}
\eeq
where $\aw (\mu)$ is given in Eq. (\ref {20}).

We note that the right-hand side of Eq.~(\ref{30}) for the kinetic 
operator was obtained, with the $\as^2$ accuracy,  
in \cite{xi} considering the zero-recoil processes, where the
dipole radiation is absent and the inelastic structure functions appear
only in the order $1/m_Q^2$. Once again, the OPE
ensures that the result is process-independent, which is demonstrated
here at the quite nontrivial level of genuine two-loop
corrections.\footnote{To first order in $\as$ it was checked in
\cite{optical}, and, for the BLM terms, in \cite{blmope} to all orders.}
The SV limit, however, is technically important since 
the separation between high- and low-energy hadronic states by means of 
the cutoff in $\omega$ is
velocity-independent only through order $\vec{v}^{\,2}$. 

The similar $\mu$-dependence holds for the renormalized slope of the
Isgur-Wise function when it is defined in the same physical way
\cite{rev}, based on the Bjorken sum rule:
\beq
\mu\frac{{\rm d}
\left [\varrho^2(\mu)-\frac{1}{4} \right ]}{{\rm d}\mu}\;=\;
\frac{4}{3}\,
\frac{2\aw(\mu)}{3\pi}\;.
\label{30a}
\eeq
Due to the short-distance effects, the observable transition form factors 
do not
have a literal heavy quark limit at $\vec v \ne 0$ but vanish, and
require factoring out the perturbative suppression due to gluon
radiation, to yield the
effective $\mu$-dependent IW function $\xi(v^2;\,\mu)$. The 
renormalization factors
generally differ depending on the way the $\mu$-cutoff is introduced in the
perturbative factors. 
In analogy with Eq.~(\ref{5}), it is convenient to define
$\xi(v^2;\,\mu)$ for small $v$ as 
\beq
\xi(v^2;\,\mu)\; = \;\lim_{m_Q\ra \infty} 
\frac{F(v)} {\left[\frac{1}{2\pi}\int_0^\mu\; w(\omega,\, \vec
v\,)\:{\rm d}\omega \right]^{1/2} }\;,\qquad 
\varrho^2(\mu) = 
-2\frac{{\rm d}\xi(v^2;\,\mu)}{{\rm d}\vec{v}^{\,2}}\vert_{v=0}\;,
\label{ximu}
\eeq
where $F$ is the form factor, for example, of 
the vector current $\bar Q \gamma_0 Q$ 
and $w$ is the
corresponding structure function. This is the scheme suggested in
\cite{rev} for which Eq.~(\ref{30a}) holds.
Various definitions using the dimensional subtraction 
schemes differ from this one; they do not have such a
direct physical meaning, however.
It is interesting to note that  the
perturbative effects in $\varrho^2(\mu)$ can be viewed as a universal
$\mu$-dependent renormalization of the tree-level constant accompanying
it ($-1/4$ for scalar mesons or $0$ for baryons).

The RG equations (\ref {30}) are directly proportional to $\aw$.
Hence, the application of the OPE shows that the 
first-order running is not renormalized in the Abelian theory (without
light fermions) to {\it all orders} in the coupling
\cite{blmope}, since in such a theory $\aw$ identically coincides with 
$\al(0)$, as discussed above.

According to \cite{five}, one can calculate perturbatively a physical
(gauge-invariant) 
low-scale running mass $m_Q(\mu)$ with $\mu \ll m_Q$ by subtracting, 
order by order, the
infrared part given by the SV sum rules, from the pole mass $m_Q^{\rm pole}$:
$$
m_Q(\mu)=\left[m_Q^{\rm pole}\right]_{\rm pert} - 
\left[ \Lam(\mu)\right]_{\rm pert} - \frac{1}{2m_Q(\mu)} 
\left[\mu_\pi^2(\mu) \right]_{\rm pert}\;.
$$
This mass determines the kinetic energy term in the renormalized heavy quark
Hamiltonian when it is expanded in the velocity of the heavy quark:
\beq
{\cal H}_Q\;=\; \bar Q A_0 Q \,+\, \frac{1}{2m_Q(\mu)} \bar
Q\left((i\vec{D}\,)^2 - c_G\frac{i}{2} \sigma G \right)Q\;+\; 
{\cal O}\left(\frac{1}{m_Q^2}, \frac{\vec v^{\,4}}{m_Q} \right) \;.
\label{kin}
\eeq
The two-loop relation between this $m_Q(\mu)$ and the 
$\overline{\rm MS}$ mass is 
\begin{eqnarray}
\label{mass}
m_Q(\mu) = {\bar m} \Bigg \{1&+&\frac {4}{3}\frac {\alpha _s (\bar m)}{\pi}
\left (1 - \frac {4}{3} \frac {\mu}{\bar m} - \frac {\mu^2}{2\bar m^2} \right)
\\
+ \left (\frac {\alpha _s (\bar m)}{\pi} \right )^2 \Bigg [
K &-&\frac {8}{3} +  \frac {\mu}{\bar m} \left (\frac {8\beta _0}{9} X_1
+ \frac {8\pi^2}{9} - \frac {52}{9} \right )
+
\frac {\mu ^2}{\bar m ^2} \left (\frac {\beta _0}{3}X_2 
+ \frac {\pi^2}{3} - \frac {23}{18} \right )
\Bigg ] \Bigg \}
\nonumber
\end{eqnarray}
where 
\beq
K = \frac{\beta_0}{2}\left(\frac {\pi^2}{6} +\frac {71}{48} \right)\,+\,
\frac {665}{144} + \frac {\pi^2}{18}\left(2\ln{2} - \frac{19}{2}\right)
-\frac {1}{6} \zeta (3), 
\eeq
\beq
X_1 = \ln{\frac {2\mu}{\bar m}} - \frac {8}{3},~~~~
X_2 = \ln{\frac {2\mu}{\bar m}} - \frac {13}{6}
\eeq
and $\bar m = \bar m(\bar m)$ is the $\overline{\rm MS}$ mass normalized at
the scale $\bar m$. 
We neglected small terms $\sim \mu^3/m_Q^2$ which can be
incorporated in the same way, if necessary. 
We used here  the 
${\cal O}(\alpha _s ^2)$ relation between the pole mass 
and the $\overline{\rm MS}$ mass \cite{Broadhurst}.

The BLM part of the $\as^3$ terms in $m_Q(\mu)/\bar{m}$ in
Eq.~(\ref{mass}) is given by
$$
\frac{m_Q(\mu)}{\bar{m}}
\;\stackrel{{\cal O}(\as^3)^{\rm BLM} }{=} 
\;
\left(\frac{\beta_0}{2}\right)^2 
\left\{\frac{2353}{2592}+\frac{13}{36}\pi^2 +\frac{7}{6} \zeta(3)-
\frac{16}{9}\frac{\mu}{m}\left[X_1^2+
\frac{67}{36}-\frac{\pi^2}{6}\right]
-
\right. 
$$
\beq
\left.
- \;
\frac{2}{3} \frac{\mu^2}{m^2} 
\left[X_2^2+
\frac{10}{9}-\frac{\pi^2}{6}\right]
\right\} 
\left(\frac{\as}{\pi}\right)^3\;,
\label{34}
\eeq
where the expressions of \cite{beta0} are used.

A word of clarification is in order regarding the definition of $\aw$. We
did not specify in Eq.~(\ref{12}) which heavy quark masses should be used
to define the reference point for $\omega$. Clearly, they must be 
low-scale masses, determined in one and 
the same way for both initial and final
state quarks. The dependence on the choice then disappears in the
difference $(m_Q-m_{\tilde Q})$ in the heavy quark limit, but still
persists in the kinetic energy of the recoiling quark given by the last
term in (\ref {12}).
 
Since a possible difference in the definition of
$\omega$ by itself is quadratic in velocity, the 
choice of 
the normalization point for the quark masses 
does not
matter here.\footnote{One should, 
actually, take the hadron masses in the kinematics.}
This is, again, the reason why the analysis becomes so much
simpler in the SV kinematics.

We conclude with addressing the question of nonperturbative
effects in the dipole radiation. When the 
effects  
suppressed by powers of
$\Lam/\omega$ are not neglected, the nonperturbative physics 
starts to 
play a role. 
In particular, the actual structure functions in
Eqs.~(\ref{9}) refer to a particular heavy hadron ($B$ meson,
$\Lambda_b$ etc.) but not an isolated heavy quark. Likewise the
complete effective coupling $\aw$ defined in this way, as any other
effective charge \cite{grunberg}, acquires certain power-suppressed
nonperturbative contribution. For example, this coupling
would differ in $B$ and
$\Lambda_b$ even in the heavy quark limit.

One can quantify the nonperturbative effects in $\aw(\omega)$
by constructing its 
$1/\omega$ expansion.\footnote{We address here only
the effects of the infrared origin and discard a more theoretical
problem of ultraviolet renormalons \cite{vz} or effects of small-size 
instantons. Thus we can merely consider, for example, the 
difference between the coupling from heavy mesons and heavy baryons.} 
We will show now that the methods used in the OPE are applicable here.

In general, the transition amplitude
$T(\omega,\,\vec{v}\,)$ has a number of cuts.
The physical `decay' cut, which generates $w(\omega, \vec v)$, 
starts at $\omega\simeq 0$. 
The other cuts go away to infinity when $m_{Q,\tilde Q}\ra \infty$ 
\cite{rev}.
Therefore, $\aw(w)$ is completely determined by the analytic
continuation of the nonrelativistic part of 
$T(\omega,\vec v)$ to positive $\omega$.
In the dispersion relation 
\beq
T(\omega,\,\vec{v}\,)\;=\; \frac{1}{2\pi}
\int_{\omega' \ge 0}\,
\frac{w(\omega',\,\vec{v}\,)}{\omega'-\omega-i\epsilon}\, d\omega' +
\frac{1}{2\pi i}\int_{\rm other\;cuts}\,
\frac{{\rm disc}\:T(\omega',\,\vec{v}\,)}{\omega'-\omega-i\epsilon}\, d\omega' 
\label{disp}
\eeq
the contribution of the additional cuts vanishes in the heavy quark
limit. At the same time, the large-$\omega$ behavior  
of $T(\omega)$ in the complex plane and, therefore, the asymptotics of 
$\aw(\omega)/\omega $
is determined by the standard OPE expansion  which expresses
the contribution of
the soft physics in terms of the expectation values of local heavy quark 
operators $\bar Q O_k Q$ of dimension $k+3$:
\beq
\frac{\delta\aw(\omega)}{\omega}\;=\;
\sum_k\; c_k \;\frac{1}{2M_{H_Q}}
\frac{\matel{H_Q}{\bar Q O_k Q}{H_Q}}{\omega^{k+1}}\;.
\label{ope}
\eeq
In this respect, the case of $\aw$ is similar to the effective
couplings defined through the Adler $D$-function $\as^{(D)}(Q^2)$ (with 
$D(Q^2)\equiv 1+\as^{(D)}(Q^2)/\pi$) or a similar $\as^{(R)}(s)$ effective
coupling which measures the strong-interaction corrections in the
absorptive part of the correlator of the vector currents of massless
quarks in the Minkowski domain. In that cases the OPE predicts
that the nonperturbative effects associated with the vacuum
expectation value of the operator
$G_{\al\be}^2$, are suppressed by a fourth power of energy.

The first nontrivial heavy quark operator is the kinetic operator $\bar
Q (i\vec D\,)^2 Q$, which, by dimensional arguments, would lead to
$1/\omega^2$ effects. However, it does not contribute 
to $\aw$. Indeed, we can consider the initial heavy quark moving 
with the spacelike momentum $|\vec{p}| \ll \omega$. The structure function
$w_{\vec{p}\,}(\omega,\,\vec{v}\,)$ for 
such a state $|Q_{\vec{p}}\rangle$ is simply 
related to the one at rest, $w_0(\omega,\,\vec{v}\,)$: 
$$
w_{\vec{p}\,}(\omega,\,\vec{v}\,)\; = 
\; w_0(\omega-\vec{v}\vec{p},\,\vec{v}\,)\; +
{\cal O}\left(1/m_Q,\,\vec{v}^{\,5}\right)\;.
$$
If we consider a superposition of such states 
$\int d^3 \vec{p}\,\Phi(\vec{p}^{\,2})\,|Q_{\vec{p}}\rangle$  
which is invariant under rotations, its  structure function 
differs from $w_0$ only in
terms $\sim \vec{p}^{\,2} \vec{v}^{\,4}$  which are negligible in the SV
limit. On the contrary, the kinetic expectation value for such a state
is non-zero and 
equals to $\aver{\vec{p}^{\,2}}$. This proves that the corresponding
coefficient function $c_2$ must vanish. 

With the heavy quark spin decoupled, only spin-singlet operators can appear 
in the expansion Eq.~(\ref{ope}). 
There is one spin-singlet, so-called Darwin operator 
$O_D= -i\bar Q D_k D_0 D_k Q$ which yields $1/\omega^3$-suppressed
terms.\footnote{At the level of
nonperturbative effects, for heavy quark states with spin of light degrees of 
freedom $j \ge 1$ there can be the second Lorentz structure in 
$d(\omega) \sim (\vec q \vec j)^2$ if averaging over these spin states is not
performed. Then an independent $D=6$ operator with 
two spacelike derivatives can 
appear.}  We do not see any reason why such terms 
cannot appear. Moreover, using the sum rule for the Darwin
term \cite{rev,motion,pirjol}, 
one finds a unique contribution associated
with its logarithmic anomalous dimension: 
\beq
\frac{\delta_{\rm D}\aw(\omega)}{\aw(\omega)} = 
\frac{3\pi\hat\gamma_D(\omega)}{8\aw(\omega)} 
\frac{\rho_D^3(\omega)}{\omega^3}
\simeq 
\left(\frac{\as(\mu)}{\as(\omega)}\right)^{\gamma_D/\beta_0}
\frac{3\gamma_D \rho_D^3(\mu)}{16\omega^3} \,\approx \,
-\left(\frac{0.55 \GeV}{\omega}\right)^3
,
\label{darwin}
\eeq
where 
$$
\hat\gamma_D(\mu) \rho_D^3(\mu) = \mu\frac{\rm d}{{\rm d} \mu} \rho_D^3(\mu) 
\;,\;\;\;  \hat\gamma_D(\mu) \simeq \gamma_D
\frac{\aw(\mu)}{2\pi}\;, \;\;\;  \gamma_D = -\frac{13}{2}\;, 
%-\beta_0+\frac{9}{2}-\frac{2}{3}n_f= -\frac{13}{2} \;,
$$
$$
\rho_D^3(\mu)\;=\; \frac{\matel{B}{O_D}{B}_\mu}{2M_B}\simeq
\frac{2\pi\as}{9} \tilde 
f_B^2 M_B \approx 0.1\GeV^3\; \mbox{ at } \; \as(\mu)\simeq
1\;.
$$
We used here the factorization estimate of $\rho_D^3$ for pseudoscalar
(vector) mesons \cite{motion};
the anomalous dimensions of the $D=6$ operators were calculated in
\cite{vslog}.
In any case, the nonperturbative effects are expected to
die out quickly with energy: 
$$
\delta_{\rm np}\aw(\omega)\; \sim \;
\left(\frac{\Lam}{\omega}\right)^k\;, \qquad k\ge 3\:.
$$

Even though $\aw$ is one of many possible effective couplings in QCD, it is 
most appropriate for heavy flavor electroweak transitions. Moreover, it has
an additional advantage compared to a few alternative ones
considered in the literature, like $\as^{(D)}$ or $\as^{(R)}$. The
latter determine 
the {\em deviation} of the corresponding observables, $D(Q^2)$ 
or $R(s)$, from their tree-level value $1$ -- whereas $\aw$ directly
measures the strength of the interaction.
This is an obvious advantage for its accurate 
determination.
In this respect, the two-loop relation for $\aw$ derived 
in the present letter,
has its counterpart in three-loop calculations of $D$-function or $R({\rm
e^+e^- \ra hadr})$. 

 From this perspective, $\aw$ is closer to the so-called $V$-scheme
coupling $\as^{(V)}$ related to the heavy quark potential $V(R)$
\cite{appel,BLM,peter}. 
The effective coupling
$\as^{(V)}$ is physically appropriate in quarkonia and near-threshold
perturbative calculations. However, its exact definition in higher orders 
is not evident 
at the moment. As a matter of fact, it requires  
knowledge 
of the heavy
quark potential at all distances $R$ including $R\ra \infty$, 
even when
$\as^{(V)}(q^2)$ is evaluated in the perturbative domain. In particular, 
appearance of
terms $\as^k\ln{\as}$ with $k\ge 4$ in its relation to the $\am$
\cite{appel}  may indicate that this is not a genuinely
short-distance quantity.  In contrast, $\aw$ is completely defined, not
only in the perturbation theory but even beyond it. This  clearly
represents a certain theoretical advantage.

One aspect of $\aw$ must be noted: it is defined in Minkowski space and
in this respect is similar to $\as^{(R)}$ but not to $\as^{(D)}$. Since,
as was mentioned, in this problem the Euclidean OPE is still applicable,
it is not too important. Nevertheless, the so-called ``exponential''
terms in $\as^{(D)}$ related to the asymptotic nature of the power
expansion of the OPE, which are truly exponentially suppressed in the
Euclidean domain, may  yield only oscillating power-suppressed component
in $\as^{(R)}$, and, similarly in $\aw$. The recent theoretical
discussion can be found in \cite{inst}. 
The coupling $\aw$ may also show
some (smeared) resonance  structure at the heavy flavor thresholds. 

The Euclidean counterpart of $\aw$
can also be defined in
a straightforward way. One can either use a Borel transform image of the
structure function $\aw(\omega)/\omega$, or usual dispersion integral.
These technical details will be given elsewhere.

The ${\overline {\rm MS}}$ 
coupling $\am$ proved to be
indispensable for complicated multi-loop calculations. On the other hand,
it is rather unphysical in some respects and its use as an
expansion parameter sometimes obscures the perturbative  
expansion \cite{stan}. At the
moment the complete ${\cal O}(\as^2)$ relation between $\am$ and $\aw$ is 
enough.  The order-$\as^3$ calculation seems feasible as
well.\footnote{It is interesting to attempt calculating the 
perturbative corrections to $\aw$ by considering dipole radiation in
the nonlinear non-Abelian classical colordynamics.}
It is natural to think that using $\aw$ or its Euclidean counterpart is
advantageous in heavy quark decays, and may improve the accuracy of the
perturbative estimates. \vspace*{.15cm}

To summarize, we derived the non-Abelian dipole radiation by
nonrelativistic color particle to second order in the strong coupling.
By virtue of the SV sum rules this determines the two-loop power mixing
of the number of spin-singlet effective local heavy quark operators. We
give the two-loop relation between the gauge-invariant short-distance 
low-scale heavy quark mass $m_Q(\mu)$ suitable for the Wilson OPE 
and the value of
$m_Q(m_Q)$ in the $\overline{\rm MS}$ scheme. We discuss the properties
of the effective dipole radiation coupling $\aw$ which is useful
for heavy quark physics. 
\vspace*{.3cm}\\
{\bf Acknowledgments:} \hspace{.4em} N.U. is grateful to
I.~Bigi, M.~Shifman and A.~Vainshtein for fruitful collaboration and to 
S.~Brodsky, Yu.~Dokshitzer, A.~Mueller and M.~Voloshin for illuminating 
discussions of perturbative aspects. 
This work was supported in part by NSF under the grant
number PHY 92-13313, by BMBF under grant number BMBF-057KA92P
and by Graduiertenkolleg ``Teilchenphysik'' at the University 
of Karlsruhe.

\end{document}